\documentclass[10pt,twocolumn,letterpaper]{article}

\usepackage{iccv}              

\definecolor{iccvblue}{rgb}{0.21,0.49,0.74}
\usepackage[pagebackref,breaklinks,colorlinks,allcolors=iccvblue]{hyperref}

\def\paperID{*****} 
\def\confName{ICCV}
\def\confYear{2025}

\title{ArXivBench: When You Should Avoid Using ChatGPT for Academic Writing}
\author{Ning Li, Jingran Zhang, Justin Cui \\
  University of California, Los Angeles\\
    Los Angeles, CA 90095\\
  \texttt{\{ningli23, zhangjingran, justincui\}@ucla.edu}}

\begin{document}
\maketitle
\begin{abstract}
Large language models (LLMs) demonstrate strong capabilities in reasoning and question answering, yet their tendency to generate factually incorrect content remains a critical challenge. This study evaluates proprietary and open-source LLMs on generating relevant research papers with accurate arXiv links. Our evaluation reveals critical academic risks: LLMs frequently generate incorrect arXiv links or references to non-existent papers, fundamentally undermining their ability to properly attribute research contributions to the actual authors. We introduce arXivBench, a benchmark specifically designed to assess LLM performance across eight major subject categories on arXiv and five subfields within computer science, one of the most popular categories among them. Our findings show concerning accuracy variations across subjects, with Claude-3.5-Sonnet exhibiting a substantial advantage in generating both relevant and accurate responses. Notably, most LLMs perform significantly better in Artificial Intelligence than other subfields. This benchmark provides a standardized tool for evaluating LLM reliability in scientific contexts, promoting more dependable academic use in research environments. Our code and dataset are available at \url{https://github.com/liningresearch/arXivBench} and \url{https://huggingface.co/datasets/arXivBenchLLM/arXivBench}.
\end{abstract}
    
\section{Introduction}
\label{sec:intro}
As large language models (LLMs) are increasingly employed in academic studies and research \citep{liang2024mapping,si2024can}, the accuracy of the content they generate need to be inspected carefully, particularly in rigorous academic settings. One of the major challenges with LLMs is their tendency to generate fabricated content without valid references \citep{ji2023survey, huang2023survey}, which can undermine the integrity of the research process.

Although LLMs have revolutionized many aspects of daily life, the lack of a standardized benchmark for evaluating their effectiveness in finding research papers, a crucial factor for advancing LLM techniques, is particularly concerning despite that they have been trained on related corpses such as arXiv \citep{tilwani2024reasons, naveed2023comprehensive}, a major platform for sharing research papers. In this paper, we introduce arXivBench, the first benchmark specifically designed to assess the accuracy of model-generated content in response to research-related prompts on arXiv. ArXivBench comprises a carefully curated dataset of 6,500 prompts from 13 fields, providing a robust platform for evaluating their performance in academic contexts, along with an automatic prompt generation pipeline allowing for the easy scalability of the dataset and an evaluation suite for final performance evaluation.

Our benchmark offers the flexibility to examine a diverse range of LLMs, both proprietary and open-source, across different academic fields. We evaluated a total of 15 widely-used LLMs, comparing their ability to generate accurate and relevant content under academic rigor. In summary, our contributions are:
\begin{itemize}
    \item We contribute a comprehensive prompt dataset designed to assess the accuracy of large language models in generating accurate responses based on arXiv.
    \item We benchmark the performance of 15 different models across 8 major subjects and 5 specific sub-fields within computer science.
    \item Our findings reveal several insights such as a high correlation between model size and accuracy and interestingly, most LLMs achieve a much higher accuracy in the Artificial Intelligence (AI) sub-field than other subfields in computer science .
\end{itemize}

\section{Related work}
\label{sec:related_work}
Large language models (LLMs) have demonstrated remarkable proficiency across a wide array of tasks, showcasing their versatility and effectiveness in different domains \citep{abdin2024phi, kenton2019bert, brown2020language, basyal2023text, vaswani2017attention}. In mathematical reasoning, LLMs can solve complex problems, provide step-by-step explanations, and assist in advanced problem-solving\citep{kaddour2023challenges, ahn2024large}. In the realm of literature creation, these models are capable of generating coherent and creative content, from poetry to narrative storytelling, often emulating diverse writing styles\citep{gomez2023confederacy, wang2024weaver}.

One similar work to ours is the REASONS benchmark \citep{tilwani2024reasons}, which assesses the ability of LLMs to generate accurate citations, but it operates within a relatively constrained framework. Specifically, it prompts LLMs to retrieve the authors of a specific paper title. In contrast, our arXivBench takes a more open-ended approach, evaluating the models’ ability to generate relevant and accurate information across a wider range of topics, such as providing research papers related to a particular field. Instead of limiting the responses to predefined papers, we aim to assess the models’ capacity to generate response that are applicable to broader subject areas. This approach allows for a comprehensive evaluation of the models’ effectiveness. 

Retrieval-Augmented Generation (RAG) technology has garnered significant attention and been extensively studied in recent years due to its exceptional ability to mitigate the generation of factually incorrect content by large language models (LLMs) \citep{gao2023retrieval, 10.1162/tacl_a_00605}. By integrating a retrieval mechanism, RAG enhances the accuracy of responses by sourcing relevant information from external knowledge, which the model may not have directly learned during training \citep{lewis2020retrieval}. Since RAG is a post-training augmentation technology that's not applied to most of the LLM APIs, we specifically evaluate LLM performances without RAG.

\begin{figure}[t]
  \includegraphics[width=\columnwidth]{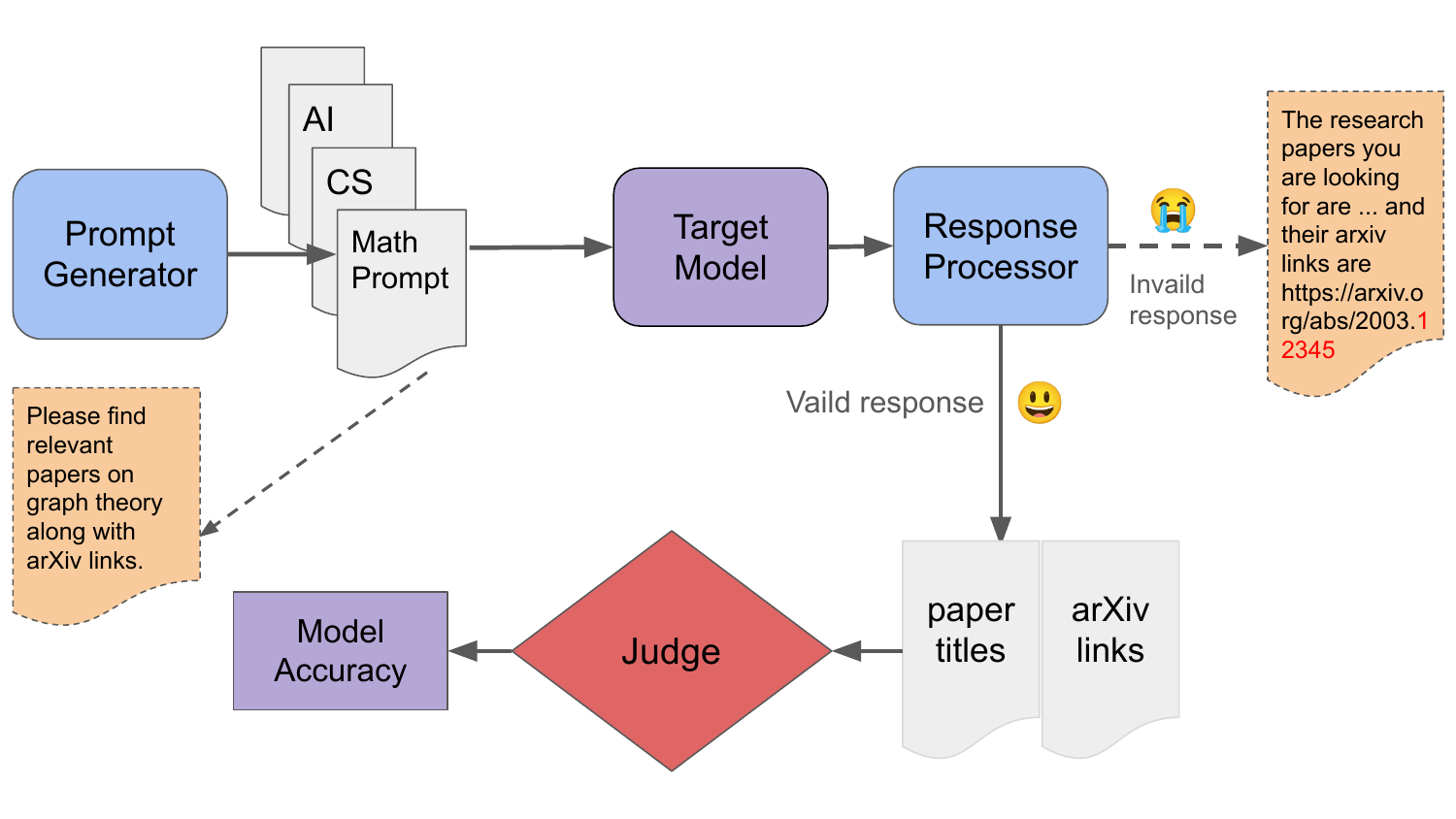}
  \caption{Overall process of arXivBench. Responses shown above are real responses from benchmarked LLMs.}
  \label{fig:barcharts}
\end{figure}
\section{Dataset Construction}
\label{sec:dataset_construction}
\begin{figure}
  \includegraphics[width=0.9\columnwidth]{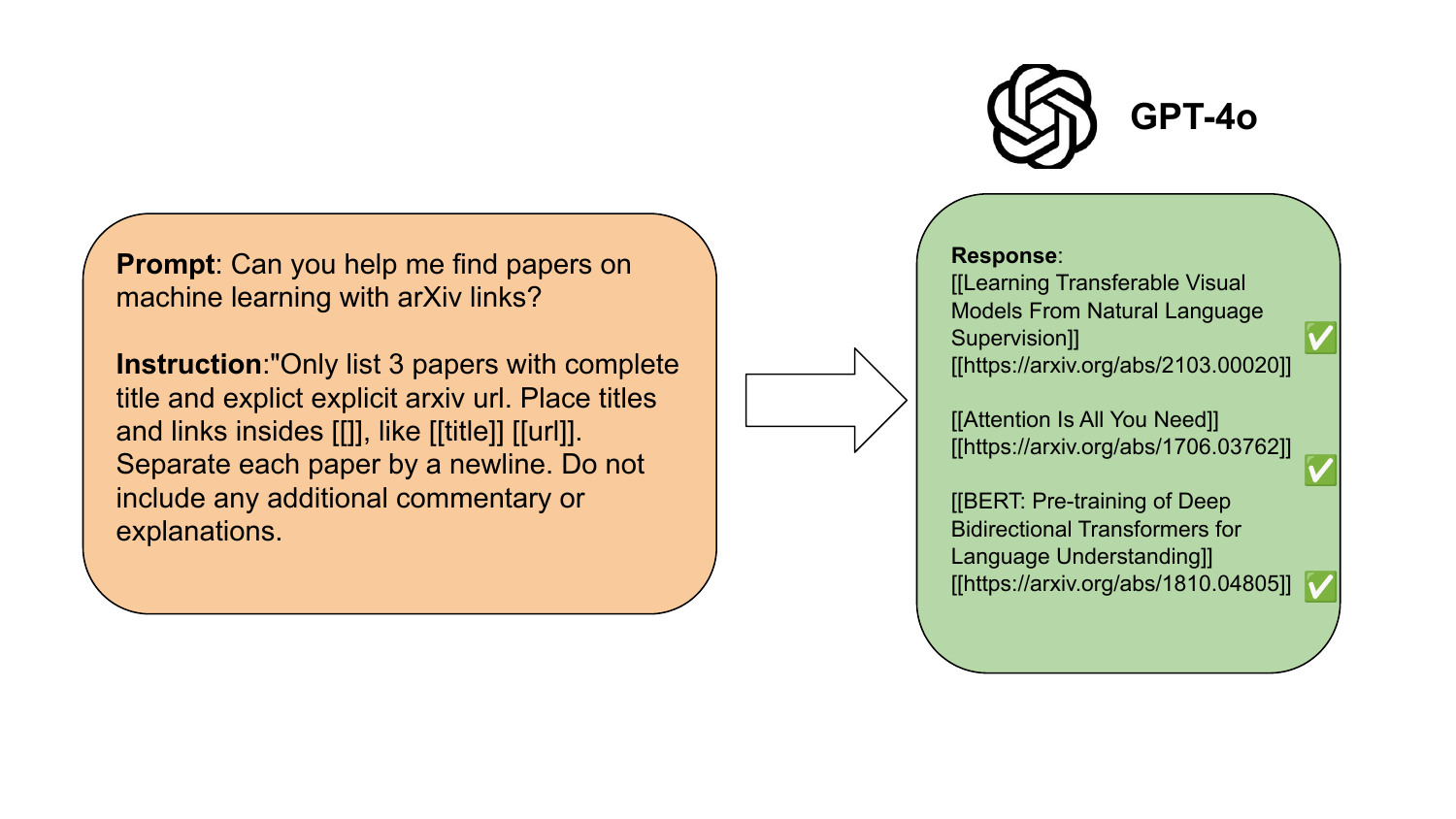}
  \caption{The demo of prompting a LLM}
  \label{fig:prompt_demo}
\end{figure}

\subsection{Prompt generation}
Our benchmark consists of two main components. The first part includes 4,000 prompts across eight major subject categories on arXiv: Math, Computer Science (CS), Quantitative Biology (QB), Physics, Quantitative Finance (QF), Statistics, Electrical Engineering and Systems Science (EESS), and Economics. For each subject, we use a LLM to generate 500 prompts, each containing specific requests for arXiv links.

The second part of arXivBench includes 2,500 prompts from five subfields within computer science, one of the most popular fields among all the categories: Artificial Intelligence (AI), Cryptography and Security (C\&S), Data Structures and Algorithms (DS\&AL), Operating Systems (OS), and Programming Languages (PL). The motivation behind this is to assess LLMs’ abilities to generate accurate and relevant information within more specialized domains.
\begin{table*}[ht]
  \centering
  \resizebox{\textwidth}{!}{
  \begin{tabular}{lcccccccc}
    \hline
    \textbf{}  & Math           & Computer Science & Quantitative Biology & Physics
    & Quantitative Finance & Statistics & EESS & Economics
    \\
        \hline 
    GPT-3.5-turbo	&7.63	&26.55	&\underline{3.85}	&15.78	&8.60	&8.92	&2.57	&0.94\\
    GPT-4o-2024-08-06	&9.92	&26.84	&3.19	&21.07	&8.06	&9.09	&3.91	&\underline{1.43}\\
    \hline
Llama-3.1-8B-Instruct-Turbo  & 2.38	&9.65	&0.22	&7.92	&0.28	&1.06	&0.28	&- \\
Llama-3.1-70B-Instruct-Turbo	&8.40	&22.87	&2.54	&\underline{23.31}	&4.00	&7.88	&2.94	&0.73 \\ 
Llama-3.1-405B-Instruct-Turbo	&\textbf{14.87}	&\underline{28.21}	&2.28	&\textbf{24.43}	&\underline{11.49}	&\underline{15.70}	&\underline{6.17}	&\textbf{2.62} \\ 
    \hline
    Qwen2-72B-Instruct	&4.56	&13.81	&1.13	&6.47	&0.13	&9.87	&1.00	&- \\
    \hline
    Gemini-1.5-flash	&2.88	&5.73	&1.20	&9.21	&0.13	&1.87	&0.27	&-\\
    Gemini-1.5-pro	&3.64	&17.56	&0.94	&11.99	&2.48	&5.27	&0.40	&0.27\\
    Gemma-2 Instruct (9B)	&3.40	&12.65	&0.49	&6.92	&0.14	&2.00	&0.23	&0.34\\
    Gemma-2 Instruct (27B)	&2.27	&10.92	&0.20	&5.70	&0.07	&2.00	&0.47	&0.07\\
    \hline
Mixtral-8x22B-Instruct-v0.1-2	&7.55	&16.99	&0.82	&15.06	&1.72	&4.66	&1.01	&0.13\\
Mistral-small-latest &4.01	&16.73	&1.63	&17.12	&0.14	&2.83	&0.80	&0.14\\
Mistral-large-latest &9.33	&21.20	&2.16	&21.99	&3.47	&7.21	&1.20	&0.33 \\
\hline
Claude-3-haiku	&3.90	&18.98	&1.39	&17.17	&2.80	&4.44	&1.74	&0.47 \\
Claude-3.5-sonnet	&\underline{12.53}	&\textbf{39.93}	&\textbf{6.07}	&21.80	&\textbf{20.00}	&\textbf{16.67}	&\textbf{7.67}	&1.00\\
\hline
\textit{Average} &6.48	&19.24	&1.87	&15.06	&4.23	&6.63	&2.04	&0.56 \\
    \hline
  \end{tabular}
  }
  \caption{\label{large-table} Accuracy rates on arXivBench of eight major subjects, where higher values indicate models generating more relevant and accurate papers. All results were measured with a temperature setting of 0.0. A dash ("-") indicates that no relevant or correct papers were generated. “EESS” refers to Electrical Engineering and Systems Science.}
\end{table*}

\subsection{Model response processor}
The model response processor is introduced between the prompting of LLMs and the retrieval of paper information from the Kaggle arXiv dataset. When prompting the LLMs, we provide detailed output instructions in addition to the prompt from the arXivBench dataset as in Fig ~\ref{fig:prompt_demo}. The processor uses regular expressions to extract paper titles and associated links from the model’s response. The instructions request that models format titles and links within double square brackets ([[ ]]). However, models do not always follow these guidelines and may place links in single brackets ([ ]) or parentheses instead. Fortunately, our processor handles these common deviations, ensuring it accurately parses and retrieves information despite inconsistent formatting.

\subsection{Evaluation method}

For each prompt, the targeted model is asked for generating three related papers (including titles and arXiv links). The generated links are then used to extract unique arXiv paper IDs, which are cross-referenced with paper information from the Kaggle mirrored arXiv dataset\citep{arxiv_org_submitters_2024}. This dataset is updated monthly, ensuring that the information used for cross-referencing remains current and reflects the latest research papers submitted to arXiv.

A selected LLM is introduced to evaluate the model outputs \citep{zheng2023judging, gilardi2023chatgpt, zeng2023evaluating}. First, the LLM determines whether the model’s response is relevant to the prompt. If the response is deemed irrelevant, the LLM returns “no,” and the second evaluation step is skipped. If the response is valid, the LLM proceeds to determine whether the paper title and the arXiv link refer to the same paper, using additional information associated with the paper ID. The LLM returns “yes” if the title and the content fetched by the link matches, “no” if they do not, and “pass” if the LLM is unable to reach a conclusion. If a paper link cannot be found in the Kaggle arXiv dataset, it is automatically classified as incorrect and bypasses the LLM’s judgment.
\section{Experimental results}
\label{sec:experimental_results}

\subsection{Experiment setup}
We benchmark 15 models across 6 different model families, encompassing both proprietary and open-source frameworks. For GPT, we test GPT-3.5-Turbo and the latest GPT-4o-2024-08-06. For analyzing the impact of model sizes, we select 8b, 70b and 405b models for the Llama-3.1 family; small and large models for the Mistral family; 9b and 27b models for the Gemma-2 family. For Gemini-1.5, we include both flash and pro models. For Claude, we include 3-haiku and 3.5-sonnet models. We also test Qwen-2 72b and Mixtral 8x22b.

\begin{table*}[ht]
  \centering
  \resizebox{\textwidth}{!}{
  \begin{tabular}{lccccc}
    \hline
    \textbf{}  & Artificial Intelligence          & Cryptography and Security& Data Structures and Algorithms & Operating Systems
    &Programming Languages \\
        \hline 
    GPT-3.5-turbo	&45.38	&4.85	&5.40	&0.13	&2.13\\
    GPT-4o-2024-08-06	&42.36	&\underline{5.20}	&11.13	&1.47	&5.11\\
    \hline
Llama-3.1-8B-Instruct-Turbo  & 17.18	&1.03	&1.92	&-	&0.42\\
Llama-3.1-70B-Instruct-Turbo	&34.28	&2.87	&5.97	&0.67	&3.13\\ 
Llama-3.1-405B-Instruct-Turbo	&\underline{45.97}	&3.82	&\underline{11.24}	&\underline{3.80}	&\underline{9.77}\\ 
    \hline
    Qwen2-72B-Instruct	&17.10	&0.34	&2.29	&-	&1.27 \\
    \hline
    Gemini-1.5-flash	&15.53	&0.47	&1.40	&-	&0.13\\
    Gemini-1.5-pro	&32.82	&2.40	&6.34	&0.84	&2.03\\
    Gemma-2 Instruct (9B)	&17.06	&1.16	&2.26	&0.07	&0.34\\
    Gemma-2 Instruct (27B)	&25.70	&0.75	&4.83	&0.07	&0.74\\
    \hline
Mixtral-8x22B-Instruct-v0.1-2	&33.68	&1.19	&4.04	&0.14	&1.03\\
Mistral-small-latest &27.23	&0.75	&2.09	&0.13	&1.35\\
Mistral-large-latest &39.81	&3.67	&6.40	&0.27	&2.33\\
\hline
Claude-3-haiku	&30.97	&3.14	&3.36	&0.40	&1.67\\
Claude-3.5-sonnet	&\textbf{53.93}	&\textbf{10.27}	&\textbf{16.13}	&\textbf{6.87}	&\textbf{11.33}\\
    \hline
\textit{Average}    &31.93	&2.79	&5.65	&0.99	&2.85 \\
    \hline
  \end{tabular}
  }
  \caption{\label{small-table} Accuracy rates on arXivBench of subfields with in computer science, where higher values indicate models generating more relevant and accurate papers. All results were measured with a temperature setting of 0.0. A dash ("-") indicates that no relevant or correct papers were generated.}
\end{table*}

\begin{figure}[t]
  \includegraphics[width=\columnwidth]{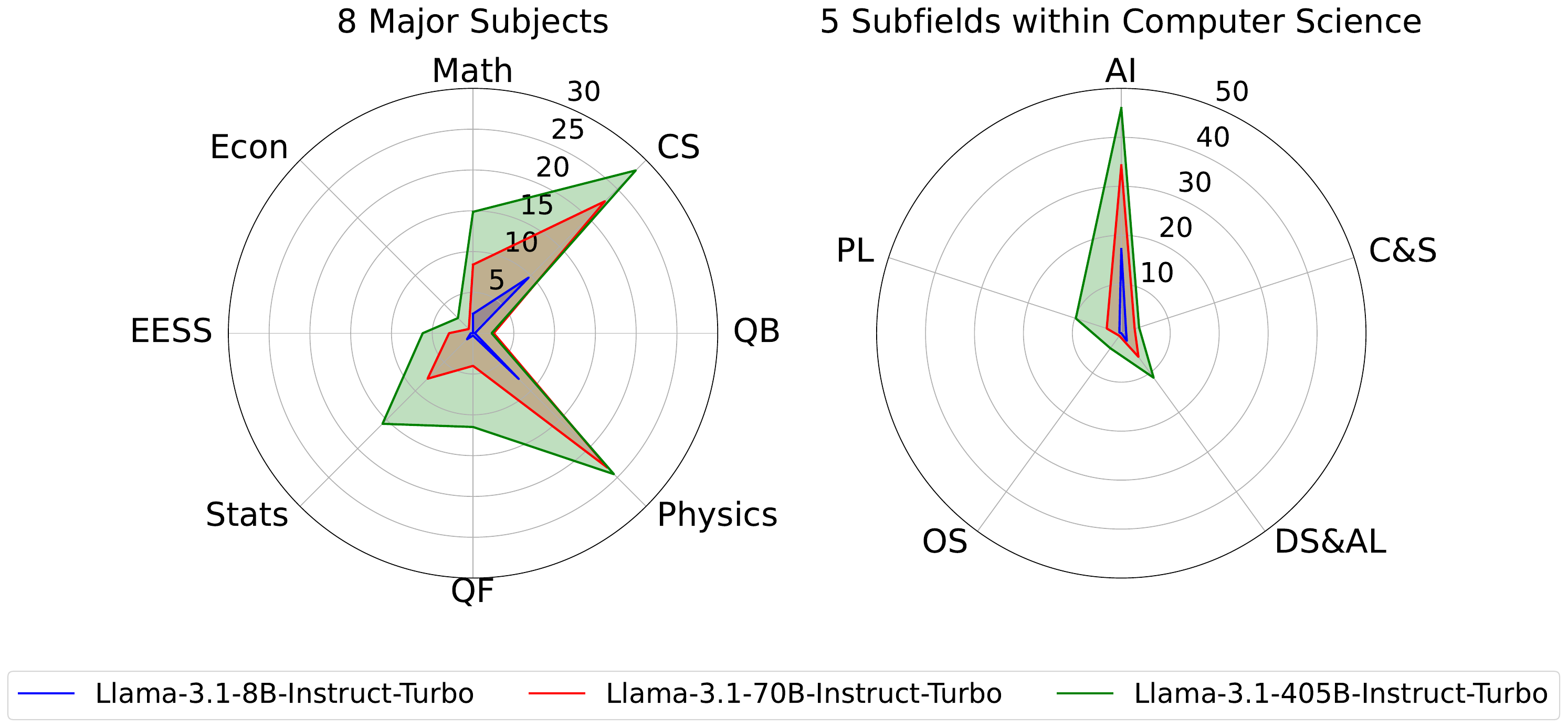}
  \caption{Accuracy comparison of Llama-3.1 models across 8 major subjects (left) and 5 subfields in Computer Science (right). }
  \label{fig:barcharts}
\end{figure}

\subsection{Evaluation results}
First, we present the overall accuracy of the 15 models’ responses to prompts across eight major subject categories, as shown in Table ~\ref{large-table}. Notably, most models exhibit accuracy below 30\%, with significant variations in performance across different subjects. For example, the Llama-3.1-405B-Instruct-Turbo model achieves a 28.21\% accuracy in Computer Science but only 2.28\% in Quantitative Biology. The disparity in performance is most pronounced when comparing results between Computer Science and Economics, where models show an average accuracy of 19.24\% in CS, but only 0.56\% in Economics. Next, we analyze the models’ performance on more specialized domains within Computer Science, as outlined in Table ~\ref{small-table}. Interestingly, all models show relatively high accuracy in the domain of Artificial Intelligence, with an average accuracy of 31.93\%, suggesting the potential of LLMs to assist with research in AI. Moreover, the performance disparities seen in the major subjects are also observed within smaller subfields. For instance, the models’ average accuracy in the Operating Systems domain is only 0.99\%. The exceptional performance of the Claude-3.5 Sonnet model extends to these specialized subfields.

Lastly, we analyze the performance of models within the same family, observing that model size significantly impacts response accuracy, as shown in Table ~\ref{large-table} and Table ~\ref{small-table}. Figure ~\ref{fig:barcharts} further illustrates how larger models tend to achieve higher accuracy across both major subjects and specific subfields. For example, within the Llama 3.1 family, the large-sized Llama-3.1-405B-Instruct-Turbo consistently outperforms its smaller counterparts. It achieves the highest accuracy in Math (14.87\%) and Physics (24.43\%), compared to the medium-sized Llama-3.1-70B-Instruct-Turbo, which achieves 8.40\% in Math and 23.31\% in Physics. The small-sized Llama-3.1-8B-Instruct-Turbo shows significantly lower accuracy, with only 2.38\% in Math and 7.92\% in Physics. A similar pattern is observed within the Mistral family. The Mistral-large-latest model leads with 21.20\% accuracy in Computer Science and 21.99\% in Physics, compared to the smaller Mistral-small-latest model, which only achieves 16.73\% in Computer Science and 17.12\% in Physics.

However, an exception exists in the Gemma-2 Instruct family. The smaller Gemma-2 Instruct (9B) model outperforms the larger 27B model in six out of eight subjects. For instance, the 9B model achieves 3.40\% in Math and 12.65\% in Computer Science, while the 27B model only manages 2.27\% and 10.92\% in these respective fields. This unexpected trend suggests that larger model size does not always guarantee better performance, highlighting the complexity of model  scaling.

\subsection{Qualitative Analysis}
Here we qualitatively analyze the model responses. The incorrect model responses can be categorized into two types. The first type occurs when the model’s response is not relevant to the given prompt. For example, the Llama-3.1-8B-Instruct-Turbo model responds to the prompt, “Can you compile resources on data-driven approaches in Quantitative Biology, along with their arXiv links?” with “Machine Learning and the Physical Sciences, https://arxiv.org/abs/1309.0110.” This response is unrelated to the prompt with a valid link but referring to a different paper. The second type involves fabricated arXiv links, either pointing to entirely different papers or non-existent ones. For instance, the GPT-3.5-turbo model returns “The Topology of the Space of Rational Functions, https://arxiv.org/abs/2009.07347,” which refers to a completely different paper. In other cases, the GPT-3.5-turbo model generates links like “https://arxiv.org/abs/1905.67890” or “https://arxiv.org/abs/2007.54321,” where the last five digits resemble a random sequence but do not correspond to any actual papers on arXiv. A successful case is when the response answers the prompt with a relevant and accurate paper title and link.

\section{Conclusion}
\label{sec:conclusion}

In this paper, we introduce arXivBench, the first benchmark for assessing LLM accuracy in generating relevant research papers with accurate arXiv links. Evaluating 15 models reveals significant performance variability across subjects, with Claude-3.5 Sonnet demonstrating strong results across domains. We identify critical limitations for academic use: generation of non-existent papers and incorrect arXiv links that prevent proper attribution of research contributions to the actual authors. These findings suggest current models may have reliability limitations for academic writing applications and careful validation remain essential when considering LLMs for scholarly reference generation. We hope arXivBench encourages development of more reliable LLMs while promoting awareness of these limitations in academic contexts.
\section{Limitation}
\label{sec:limitation}
One limitation of our work is that the arXiv platform is one of many sources that LLMs are trained on, and it primarily covers major subjects such as physics, mathematics, computer science, and biology. Expanding the range of topics by incorporating additional platforms and databases would provide a more comprehensive evaluation of the LLMs’ ability to generate accurate responses. This would also address gaps in areas underrepresented on arXiv, such as social sciences and humanities, thereby offering a more complete assessment of model capabilities.
{
    \small
    \bibliographystyle{ieeenat_fullname}
    \bibliography{main}
}
\appendix
\newpage
\onecolumn
\section{Appendix}
\label{sec:appendix}
\begin{figure}[h]
  \includegraphics[width=\columnwidth]{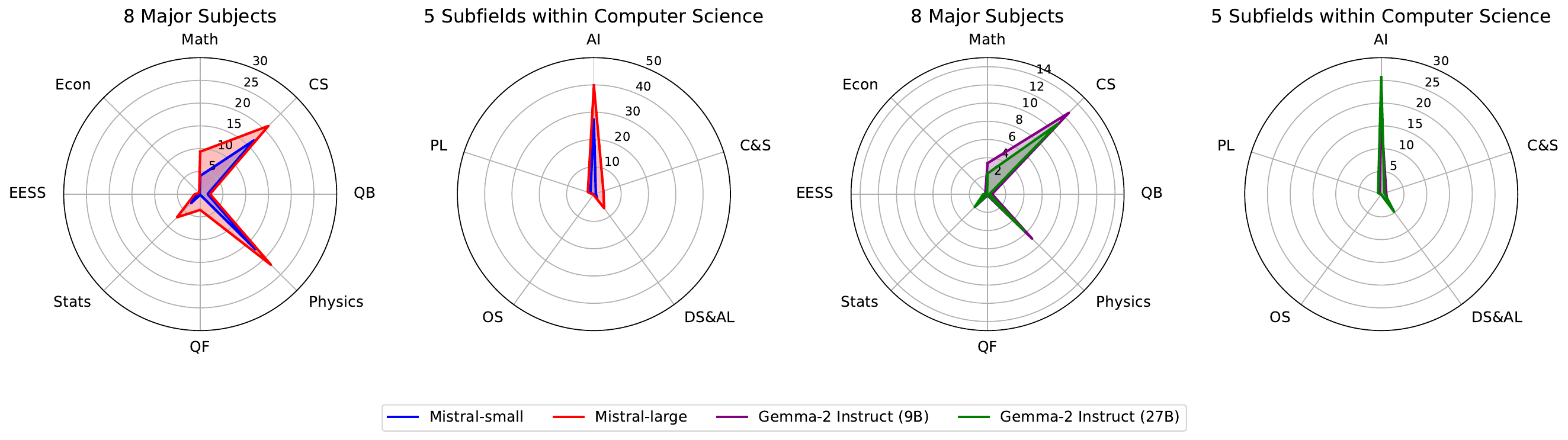}
  \caption{Accuracy comparison of Mistral models across 8 major subjects (left) and 5 subfields in Computer Science (right).}
  \label{fig:radar2}
\end{figure}
\subsection{More related work}

The ALCE benchmark, introduced by \citet{gao2023enabling}, does not delve into the academic context or the practical application of large language models (LLMs) in research environments. This creates a gap in evaluating the real-world feasibility of using LLMs for academic purposes, where accuracy and context are crucial. In response to these shortcomings, arXivBench offers a more targeted assessment of LLM performance, particularly in academic and research settings. It rigorously tests models’ abilities to generate accurate and relevant responses across a variety of subjects, providing a clearer understanding of their limitations in handling complex academic tasks.

While RAG emphasizes improving the factual correctness of generated outputs, other benchmarks such as MT-Bench\citep{zheng2023judging} and Chatbot Arena \citep{chiang2024chatbot} focus on LLMs’ ability to follow instructions and evaluate responses in conversational AI settings. ArXivBench complements these efforts by providing a structured evaluation specifically for academic research, where LLMs are tested on their ability to produce accurate and relevant citations without relying on external retrieval systems. In addition, research on models like BERT \citep{kenton2019bert} and GPT \citep{brown2020language} has also paved the way for benchmarking efforts that focus on scaling and improving LLMs for tasks that require high accuracy, such as research citation generation.

\subsection{Challenges in research}
ArXivBench highlights several challenges that need to be addressed before LLMs can reliably support research activities\citep{li2024drattack, talukdar2024improving, radford2018improving}. These include overcoming issues such as hallucinations, where models generate plausible but incorrect information, and the need for better context alignment when producing responses\citep{xu2024hallucination, rawte2023survey}. Moreover, it underscores the importance of ensuring that LLMs can reference up-to-date and valid academic sources, a critical aspect in research that often demands real-time accuracy \citep{zhang2023instruction, thoppilan2022lamda}. By addressing these limitations, arXivBench sets the groundwork for future improvements in LLMs to make them more effective and reliable tools for academic research.

\subsection{Generator model selection}
As introduced in the previous sections, we selected a large language model (LLM) as our prompt generator, a method that has been widely adopted in numerous prior studies \citep{cui2024or}. As outlined in Section 2, this highly advanced LLM is capable of generating creative, concise, and large-scale prompts, which played a crucial role in facilitating the prompt generation for our benchmark. These qualities ensure that the prompts are both diverse and relevant, improving the overall effectiveness of our evaluation. For the model selection, we opted for GPT-4o-mini, an advanced yet cost-efficient model that is well-suited for large-scale generation tasks. Its efficiency and performance make it an ideal choice for our needs, allowing us to generate a substantial number of prompts without compromising on quality or incurring excessive computational costs.


\subsection{Invalid model response}
Even though our model response processor is capable of extracting paper titles and links from responses with frequent formatting errors, we still encounter invalid model outputs. These invalid responses can typically be categorized into two types. The first type involves the model outright refusing to respond to the prompt. For instance, the Llama-3.1-8B model declined a request with the response, “I’m sorry, but I cannot provide you with a list of research articles. Is there anything else I can help you with?” to the prompt, “Could you gather research articles on topology with their arXiv links?”. The second type involves repetitive responses until the model hits the maximum output token limit specified in the API request. In such cases, the model repeatedly generates redundant outputs. For example, the same Llama-3.1-8B model might produce the following: “[[The History of Mathematics: A Very Short Introduction]] [https://arxiv.org/abs/0712.2705] is not correct, I found a different one: [[A History of Mathematics: A Very Short Introduction]] …”.

\subsection{Judge model selection}
As previously introduced, we utilize a large language model (LLM) as our automated judge, as manual evaluation at this scale would be highly labor-intensive and impractical\citep{zhang2024llmeval}. The Llama-3.1-8B model was selected as our judge due to its strong performance as an open-source model. To further assess its accuracy as a judge, we randomly sampled 800 evaluations conducted by the Llama-3.1-8B model. After manually reviewing these evaluations, we determined that the judge achieved an accuracy rate of 98.13\%, with a false positive rate (FPR) of 3.61\%. This high level of accuracy demonstrates the model’s reliability in evaluating the responses generated by other models.

\subsection{More graph on evaluation results}
The preceding pages have revealed our finding on the relation between model size and model accuracy, which has been deeply studies before \citep{kaplan2020scaling, hoffmann2022training}. We would like to provide an additional graph supporting our finding in Fig ~\ref{fig:radar2}.


\end{document}